\documentclass{article} 
\usepackage{nips10submit_e,times}

\usepackage[dvips]{graphics}
\usepackage{epsfig}
\usepackage{amsmath}

\title{Attractor Dynamics with Synaptic Depression}

\author{
C. C. Alan Fung, ~~~~~K. Y. Michael Wong\\
Hong Kong University of Science and Technology, Hong Kong, China \\
\texttt{alanfung@ust.hk, ~phkywong@ust.hk} \\
\AND
He Wang\\
Tsinghua University, Beijing, China\\
\texttt{wanghe07@mails.tsinghua.edu.cn} \\
\And
Si Wu \\
Institute of Neuroscience,\\
Chinese Academy of Sciences, Shanghai, China\\
\texttt{siwu@ion.ac.cn}
}

\nipsfinalcopy 

\begin{document}

\maketitle

\begin{abstract}
Neuronal connection weights exhibit short-term depression (STD). 
The present study investigates the impact of STD on the dynamics of a
continuous attractor neural network (CANN) 
and its potential roles in neural information processing. 
We find that the network with STD can
generate both static and traveling bumps, 
and STD enhances the performance of the network 
in tracking external inputs. 
In particular, we find
that STD endows the network with slow-decaying plateau behaviors, 
namely, the network being initially stimulated to an active state will decay
to silence very slowly in the time scale of STD 
rather than that of neural signaling. 
We argue that this provides a mechanism for neural
systems to hold short-term memory easily 
and shut off persistent activities naturally.
\end{abstract}

\section{Introduction}
Networks of various types, formed by a large number of neurons through synapses, are the substrate of brain functions. The network structure is
the key that determines the responsive behaviors of a network to external inputs, and hence the computations implemented by the neural system.
Understanding the relationship between the structure of a neural network and the function it can achieve is at the core of using mathematical
models for elucidating brain functions.

In the conventional modeling of neuronal networks, it is often assumed that the connection weights between neurons, which model the efficacy of
the activities of pre-synaptic neurons on modulating the states of post-synaptic neurons, are constants, or vary only in long-time scales when
learning occurs. However, experimental data has consistently revealed that neuronal connection weights change in short time scales, varying
from hundreds to thousands of milliseconds (see, e.g., \cite{Markram98}). This is called short-term plasticity (STP). A predominant type of STP
is short-term depression (STD), which decreases the connection efficacy when a pre-synaptic neuron fires. The physiological process underlying
STD is the depletion of available resources when signals are transmitted from a pre-synaptic neuron to the post-synaptic one.

Is STD simply a by-product of the biophysical process of neural signaling? Experimental and theoretical studies have suggested that this is
unlikely to be the case. Instead, STD can play very active roles in neural computation. For instance, it was found that STD can achieve gain
control in regulating neural responses to external inputs, realizing Weber's law~\cite{Tsodyks97,Abbott97}. 
Another example is that STD enables a network to
generate transient synchronized population firing, appealing for detecting subtle changes in the environment~\cite{Tsodyks00,Loebel02}. The STD of a neuron is also thought to play a role in estimating the information of the pre-synaptic membrane potential from the spikes it receives
\cite{Pfister09}. From
the computational point of view, the time scale of STD resides between fast neural signaling (in the order of milliseconds) and slow learning
(in the order of minutes or above), which is the time order of many important temporal operations occurring in our daily life, such as working
memory. Thus, STD may serve as a substrate for neural systems to manipulate temporal information in the relevant time scales.

In this study, we will further explore the potential role of STD in neural information processing, an issue of fundamental importance but has
not been adequately investigated so far. We will use continuous attractor neural networks (CANNs) as working models. CANNs are a type of
recurrent networks which hold a continuous family of localized active states~\cite{Amari77}. Neutral stability is a key advantage of CANNs,
which enables neural systems to update memory states or to track time-varying stimuli smoothly. CANNs have been successfully applied to
describe the retaining of short-term memory, and the encoding of continuous features, such as the orientation, the head direction and the
spatial location of objects, 
in neural systems~\cite{Ben-Yishai95,Zhang96,Samsonovich97}. 
CANNs are also shown to provide a framework for
implementing population decoding efficiently~\cite{Deneve99}.

We analyze the dynamics of a CANN with STD included, 
and find that apart from the static bump states, 
the network can also hold moving bump solutions. 
This finding agrees with the results 
reported in the literature~\cite{York09,Kilpatrick10}. 
In particular, we find that with STD,
the network can have slow-decaying plateau states, that is, the network being stimulated to an active state by a transient input will decay to
silence very slowly in the time order of STD rather than that of neural signaling. This is a very interesting property. It implies that STD can
provide a mechanism for neural systems 
to generate short-term memory and shut off activities naturally. 
We also find that STD retains the neutral
stability of the CANN, and enhances the tracking performance of the network to external inputs.

\section{The Model}
Let us consider a one-dimensional continuous stimulus $x$
encoded by an ensemble of neurons.
For example, the stimulus may represent the moving direction,
the orientation or a general continuous feature of objects
extracted by the neural system.

Let $u(x,t)$ be the synaptic input at time $t$ to the neurons
whose preferred stimulus is $x$.
The range of the possible values of the stimulus is $-L/2<x\le L/2$
and $u(x,t)$ is periodic, i.e., $u(x+L)=u(x)$. The dynamics is particularly
convenient to analyze in the limit that the interaction range $a$ is much
less than the stimulus range $L$, so that we can effectively take
$x\in(-\infty,\infty)$.  The dynamics of $u(x,t)$ is determined by the
external input $I_{\rm ext}(x,t)$, the network input from other neurons,
and its own relaxation. It is given by
\begin{equation}
    \tau_s\frac{\partial u(x,t)}{\partial t}
    =I_{\rm ext}(x,t)+\rho\int^\infty_{-\infty} dx'J(x,x')p(x',t)r(x',t)
    -u(x,t),
\label{eq:dyn}
\end{equation}
where $\tau_s$ is the synaptical transmission delay, which is typically in the order of 2 to 5 ms. $J(x,x')$ is the base neural interaction
from $x'$ to $x$. $r(x,t)$ is the firing rate of neurons. It increases with the synaptic input, but saturates in the presence of a global
activity-dependent inhibition. A solvable model that captures these features is given by $r(x,t)=u(x,t)^2/[1+k\rho\int_{-\infty}^{\infty}
dx'u(x',t)^2]$, where $\rho$ is the neural density, and $k$ is a positive constant controlling the strength of global inhibition. The global
inhibition can be generated by shunting inhibition~\cite{Hao09}.

The key character of CANNs is the translational invariance
of their neural interactions.
In our solvable model, we choose Gaussian interactions
with a range $a$, namely, $
    J(x,x')=J_0\exp[-(x-x')^2/2a^2]/(a\sqrt{2\pi})$,
where $J_0$ is a constant.

The STD coefficient $p(x,t)$ in Eq.~(\ref{eq:dyn}) takes into account the pre-synaptic STD. It has the maximum value of 1, and decreases with
the firing rate of the neuron \cite{Tsodyks98,Zucker02}. Its dynamics is given by
\begin{equation}
        \tau_d\frac{\partial p(x,t)}{\partial t} =1-p(x,t)- p(x,t)\tau_{d}\beta r(x,t), \label{eq:dp}
\end{equation}
where $\tau_{d}$ is the time constant for synaptic depression,
and the parameter $\beta$ controls the depression effect due to neural firing.

The network dynamics is governed by two time scales. 
The time constants of STD is typically
in the range of hundreds to thousands of milliseconds,
much larger than that of neural
signaling, i.e., $\tau_d\gg \tau_s$. 
The interplay between the fast and slow dynamics 
causes the network to exhibit interesting dynamical behaviors.

\begin{figure}[ht]
\centering
\begin{minipage}{16pc}
\centering
\includegraphics[width=11pc]{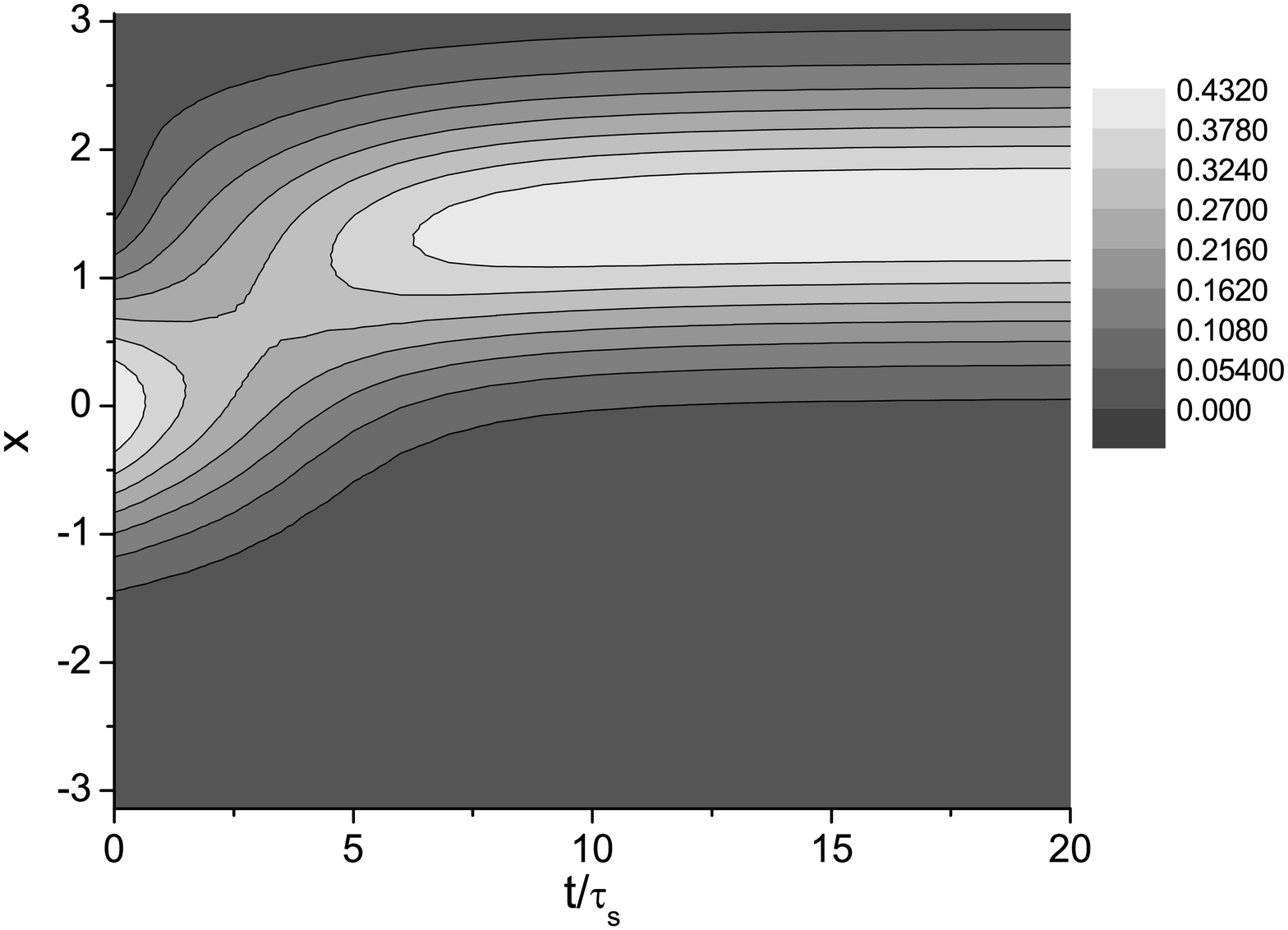}
\caption{
The neural response profile tracks
the change of position of the external stimulus 
from $z_0=0$ to $1.5$ at $t=0$.
Parameters: $a = 0.5$,
$\overline{k}=0.95$, $\overline\beta=0$, $\alpha =0.5$.
}
\end{minipage}\hspace{1pc}%
\begin{minipage}{16pc}
\centering
\includegraphics[width=12pc]{track_profile.eps}
\caption{
The profile of $u(x,t)$ at $t/\tau=0,1,2,\cdots,10$ 
during the tracking process in Fig.~1.
}
\end{minipage}
\end{figure}

\subsection{Dynamics of CANN without Dynamical Synapses}

It is instructive to first consider the network dynamics when no dynamical
synapses are included.
This is done by setting $\beta=0$ in Eq. (\ref{eq:dp}),
so that $p(x,t)=1$ for all $t$. In this case, the network can support
a continuous family of stationary states when the global inhibition
is not too strong.

Specifically, the steady state solution to Eq. (\ref{eq:dyn}) is
\begin{equation}
\tilde{u}(x|z)=u_{0}\exp\left[-\frac{(x-z)^{2}}{4a^{2}}\right]\textrm{, }\quad\tilde{r}(x|z)=r_{0}\exp\left[-\frac{(x-z)^{2}}{2a^{2}}\right],
\end{equation}
where $u_{0}=[1+(1-k/k_{c})^{1/2}]J_{0}/(4ak\sqrt{\pi})$, $r_{0}=[1+(1-k/k_{c})^{1/2}]/(2ak\rho\sqrt{2\pi})$ and $k_{c}=\rho
J_{0}^{2}/(8a\sqrt{2\pi})$. 
These stationary states are translationally invariant among themselves and have the Gaussian
shape with a free parameter $z$ representing the position of the Gaussian bumps. They exist for $0<k<k_{c}$, $k_{c}$ is thus the critical
inhibition strength.

Fung et al \cite{Fung10} considered the perturbations of the Gaussian
states. They found various distortion modes, each characterized by
an eigenvalue representing its rate of evolution in time. A key property
they found is that the translational mode has a zero eigenvalue, and
all other distortion modes have negative eigenvalues for $k<k_{c}$.
This implies that the Gaussian bumps are able to track changes in
the position of the external stimuli by continuously shifting the
position of the bumps, with other distortion modes affecting the tracking
process only in the transients.

An example of the tracking process is shown in Figs. 1 and 2, when an external
stimulus with a Gaussian profile is initially centered at $z=0$,
pinning the center of a Gaussian neuronal response at the same position.
At time $t=0$, the stimulus shifts its center from $z=0$ to $z=1.5$
abruptly. The bump moves towards the new stimulus
position, and catches up with the stimulus change after a time duration.
which is referred to as the reaction time.

\section{Dynamics of CANN with Synaptic Depression}

For clarity, we will first summarize the main results obtained on the network dynamics due to STD, and then present the theoretical analysis
in Sec.~4.

\subsection{The Phase Diagram}

In the presence of STD, CANNs exhibit new interesting dynamical behaviors.
Apart from the static bump state, the network also supports moving
bump states. To construct a phase diagram mapping these behaviors,
we first consider how the global inhibition $k$ and the synaptic
depression $\beta$ scale with other parameters. In the steady state solution of Eq. (\ref{eq:dyn}),
$u_0$ and $\rho J_0 u_0^2$ should have the same dimension; so are $1-p(x,t)$
and $\tau_d \beta u_0$ in Eq. (\ref{eq:dp}).
Hence we introduce the dimensionless parameters $\overline{k}\equiv k/k_{c}$
and $\overline{\beta}\equiv\tau_{d}\beta/(\rho^{2}J_{0}^{2})$. The
phase diagram obtained by numerical solutions to the network dynamics
is shown in Fig. 3.

\begin{figure}[ht]
\centering
\vspace{1.1pc}
\includegraphics[width=16pc]{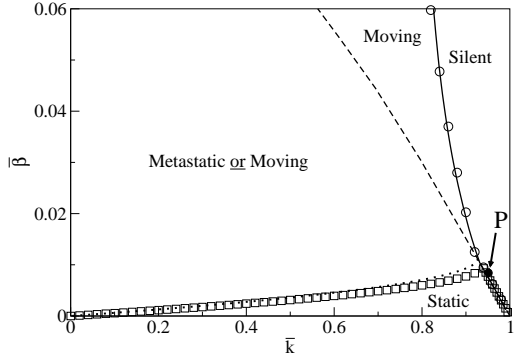}\hspace{0.5pc}%
\begin{minipage}[b]{13pc}\caption{
Phase diagram of the network states. Symbols: numerical solutions.
Dashed line: Eq.~(\ref{eq:static}).
Dotted line: Eq.~(\ref{eq:moving}).
Solid line: Gaussian approximation
using $11^{\rm th}$ order perturbation of the STD coefficient.
Point P: the working point for Figs. 4 and 7. 
Parameters: $\tau_d/\tau_s=50$, $a = 0.5/6$,
range of the network = $[-\pi, \pi)$.}
\end{minipage}
\end{figure}


We first note that the synaptic depression and the global inhibition
plays the same role in reducing the amplitude of the bump states.
This can be seen from the steady state solution of $u(x,t)$, which
reads
\begin{equation}
u(x)=\int dx'\frac{\rho J(x-x')u(x')^{2}}{1+k\rho\int dx''u(x'')^{2}+\tau_{d}\beta u(x')^{2}}.
\end{equation}
The third term in the denominator of the integrand arises from STD,
and plays the role of a local inhibition that is strongest where the
neurons are most active. Hence we see that the silent state with $u(x,t)=0$
is the only stable state when either $\overline{k}$ or $\overline{\beta}$
is large.

When STD is weak, the network behaves similarly with CANNs without STD, that is, the static bump state is present up to $\overline{k}$ near 1.
However, when $\overline{\beta}$ increases, a state with the bump spontaneously moving at a constant velocity comes into existence. Such moving
states have been predicted in CANNs \cite{York09, Kilpatrick10}, and can be associated with traveling wave behaviors widely observed in the
neocortex~\cite{Wu08}. At an intermediate range of $\overline{\beta}$, both the static and moving states coexist, and the final state of the
network depends on the initial condition. When $\overline{\beta}$ increases further, only the moving state is present.

\subsection{The Plateau Behavior}

The network dynamics displays a very interesting behavior in the parameter
regime when the static bump solution just loses its stability. In this
regime, an initially activated network state decays very slowly to
silence, in the time order of $\tau_{d}$. Hence, although the bump
state eventually decays to the silent state, it goes through a plateau
region of a slowly decaying amplitude, as shown in Fig. 4.

\begin{figure}[ht]
\centering
\vspace{3mm}
\includegraphics[width=18pc]{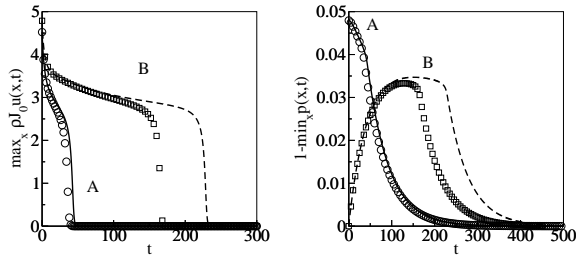}\hspace{0.5pc}%
\begin{minipage}[b]{13pc}\caption{
Magnitudes of rescaled neuronal input $\rho J_0u(x,t)$ and synaptic depression $1-p(x,t)$
at $(\overline{k},\overline{\beta}) = (0.95,0.0085)$ 
(point P in Fig. 3) and 
for initial conditions of types A and B in Fig. 8.
Symbols: numerical solutions.
Lines: Gaussian approximation using Eqs.~(\ref{eq:dU0_app})
and (\ref{eq:dp0_app}).
Other parameters: $\tau_d/\tau_s=50$, $a = 0.5$
and $x \in [-\pi,\pi)$.}
\end{minipage}
\end{figure}


\subsection{Enhanced Tracking Performance}

The responses of CANNs with STD to an abrupt change of stimulus are
illustrated in Fig. 5. Compared with networks without STD, we
find that the bump shifts to the new position faster. The extent of improvement
in the presence of STD is quantified in Fig. 6.  However, when
$\overline{\beta}$ is too strong, the bump tends to overshoot the
target before eventually approaching it.

\begin{figure}[ht]
\centering
\vspace{5mm}
\begin{minipage}{19pc}
\centering
\includegraphics[width=12pc]{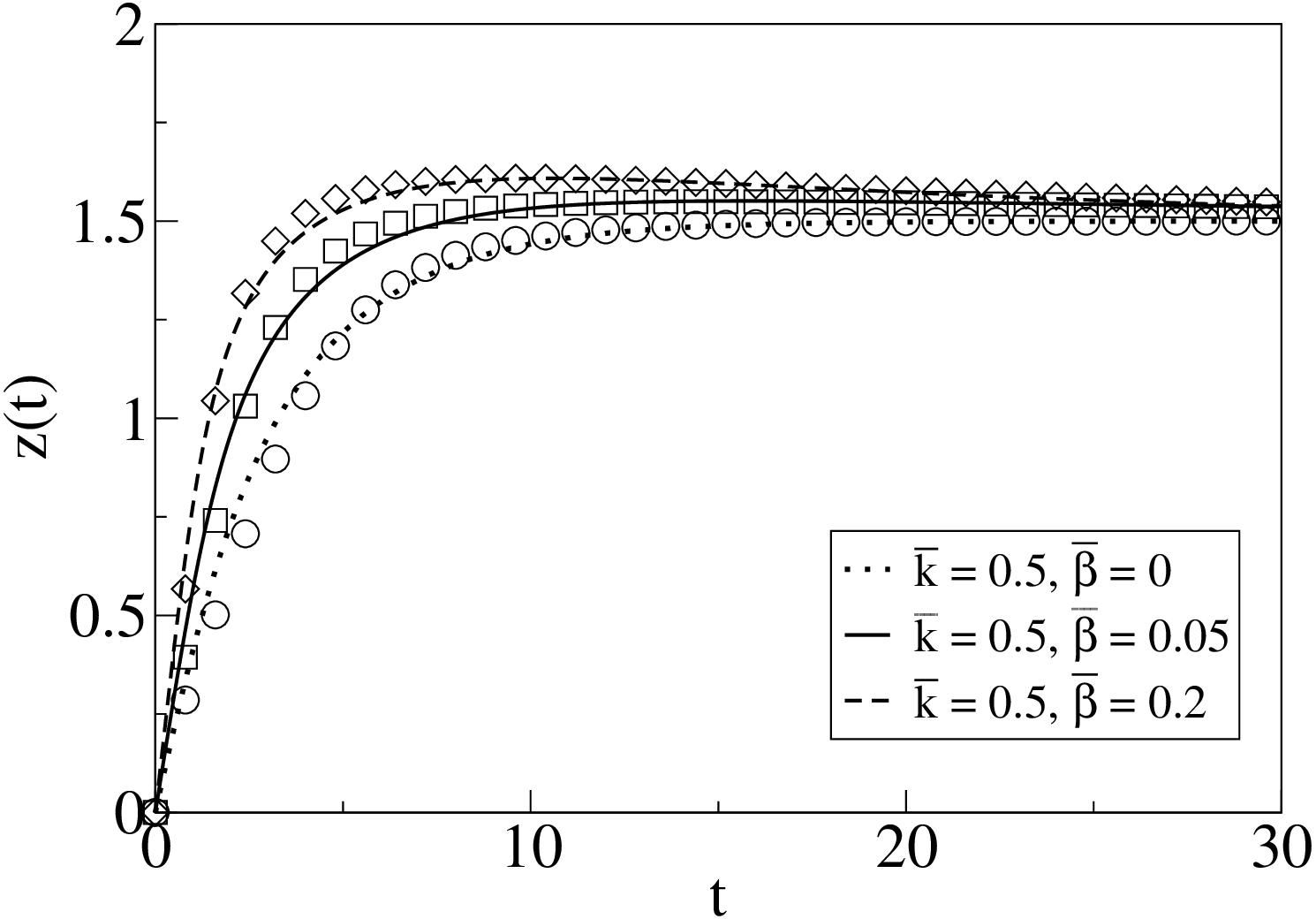}
\caption{
The response of CANNs with STD to an abruptly changed stimulus 
from $z_0=0$ to  $z_0=1.5$ at $t=0$.
Symbols: numerical solutions.
Lines: Gaussian approximation using $11^{\rm th}$ order perturbation
of the STD coefficent.
Parameters: $\tau_d/\tau_s=50$, $\alpha=0.5$, $a = 0.5$
and $x \in [-\pi,\pi)$.
}
\end{minipage}\hspace{1pc}%
\begin{minipage}{12pc}
\centering
\includegraphics[width=12pc]{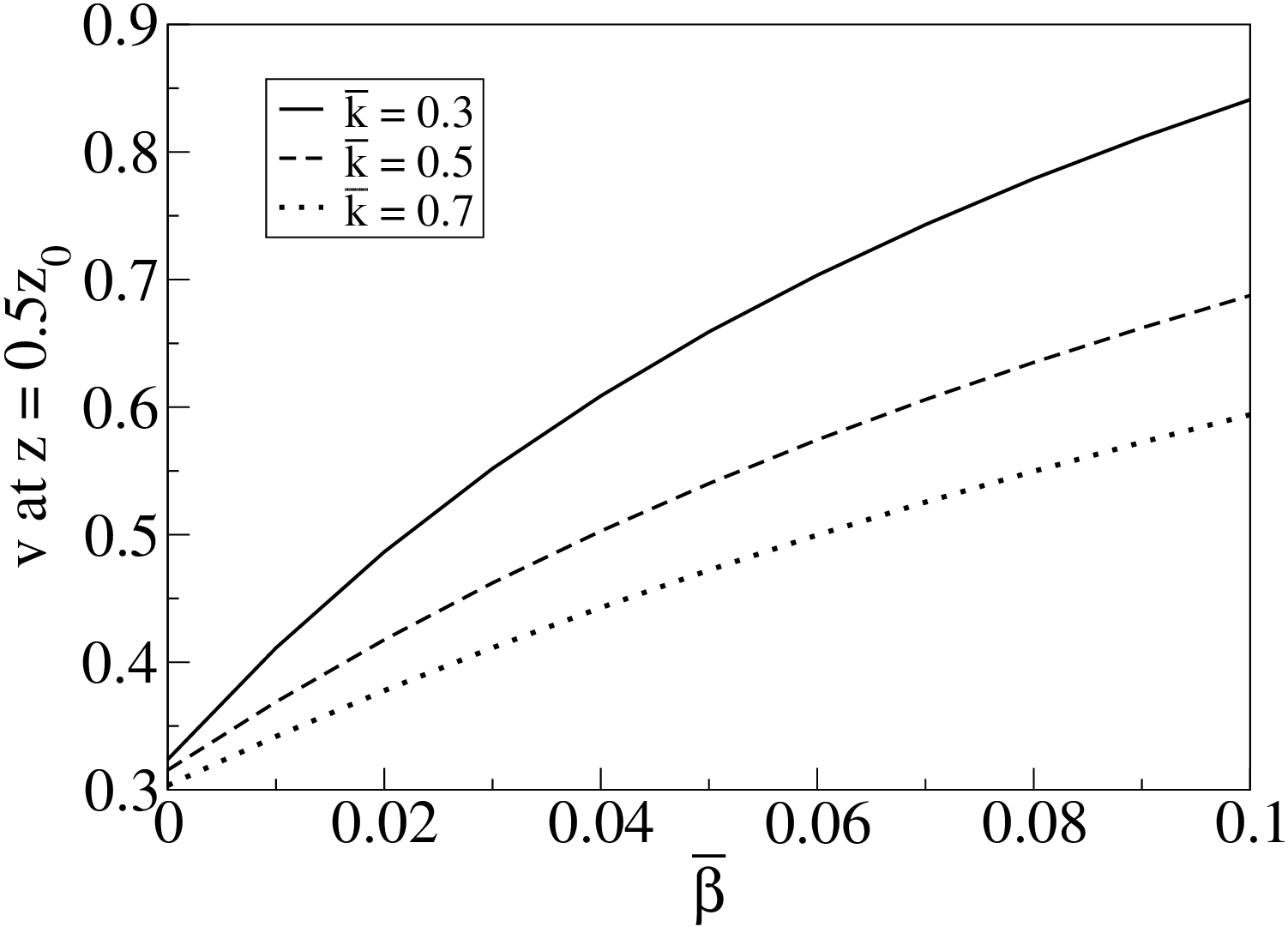}
\caption{
Tracking speed of the bump at $0.5z_0$, where $z_0$ is fixed to be 1.5
}
\end{minipage}
\end{figure}


\section{Analysis}

Despite the apparently complex behaviors of CANNs with STD, we will
show in this section that a Gaussian approximation can reproduce the
behaviors and facilitate the interpretation of the results. 
Details are explained in Supplementary Information. We observe that the 
profile of the bump remains effectively Gaussian in the presence
of synaptic depression. On the other hand, there is a considerable
distortion of the profile of the synaptic depression, when STD is strong.
Yet, to the lowest order approximation, let us approximate the profile
of the synaptic depression to be a Gaussian as well, which is valid
when STD is weak,
as shown in Fig. 7(a).
Hence, for $a\ll L$, we propose the following
ansatz\begin{eqnarray}
u(x,t) & = & u_{0}(t)\exp\left[-\frac{(x-z)^{2}}{4a^{2}}\right],\label{eq:U0_app0}\\
p(x,t) & = & 1-p_{0}(t)\exp\left[-\frac{(x-z)^{2}}{2a^{2}}\right].\label{eq:p_app0}\end{eqnarray}
When these expressions are substituted into the dynamical equations
(\ref{eq:dyn}) and (\ref{eq:dp}), other functions $f(x)$ of $x$ appear. 
To maintain consistency
with the Gaussian approximation, these functions will be approximated
by their projections onto the Gaussian functions. In Eq. (\ref{eq:dyn}), we approximate
\begin{equation}
f(x)\approx\left[\int\frac{dx'}{\sqrt{2\pi a^{2}}}f(x')e^{-\frac{(x'-z)^{2}}{4a^{2}}}\right]e^{-\frac{(x-z)^{2}}{4a^{2}}}.\end{equation}
Similarly, in Eq. (\ref{eq:dp}), we approximate $f(x)$ by its projection onto
$\exp\left[-(x-z)^{2}/(2a^{2})\right]$.

\subsection{The Solution of the Static Bumps}

Without loss of generality, we let $z=0$.
Substituting Eq. (\ref{eq:U0_app0}) and
(\ref{eq:p_app0}) into Eqs. (\ref{eq:dyn}) and (\ref{eq:dp}),
and letting $\overline{u}(t)\equiv\rho J_0 u_0(t)$, we get
\begin{eqnarray}
    \tau_{s}\frac{d\overline{u}(t)}{dt} & = &
    \frac{\overline{u}(t)^{2}}{\sqrt{2}(1+\overline{k}\overline{u}(t)^{2}/8)}
    \left[1-\sqrt{\frac{4}{7}}p_{0}(t)\right]-\overline{u}(t),
\label{eq:dU0_app}\\
    \tau_{d}\frac{dp_{0}(t)}{dt} & = &
    \frac{\overline{\beta}\overline{u}(t)^{2}}
    {1+\overline{k}\overline{u}(t)^{2}/8}
    \left[1-\sqrt{\frac{2}{3}}p_{0}(t)\right]-p_{0}(t).
\label{eq:dp0_app}
\end{eqnarray}
By considering the steady state solution of $\overline{u}$ and $p_{0}$
and their stability against fluctuations of $\overline{u}$ and $p_{0}$,
we find that stable solutions exist when
\begin{equation}
    \overline{\beta}\leq
    \frac{p_{0}(1-\sqrt{4/7}p_{0})^{2}}{4(1-\sqrt{2/3}p_{0})}
    \left[1+\frac{\tau_{s}}{\tau_{d}(1-\sqrt{2/3}p_{0})}\right],
\label{eq:static}
\end{equation}
when $p_{0}$ is the steady state solution of Eqs. (\ref{eq:dyn})
and (\ref{eq:dp}). The boundary of this region is shown as a
dashed line in Fig. 3. Unfortunately, this line is not easily observed
in numerical solutions since the static bump is unstable against fluctuations
that are asymmetric with respect to its central position. Although
the bump is stable against symmetric fluctuations, asymmetric fluctuations
can displace its position and eventually convert it to a moving bump.

\subsection{The Solution of the Moving Bumps}

As shown in Fig. 7(b), the profile of a moving bump is characterized
by a lag of the synaptic depression behind the moving bump. This is
because neurons tend to be less active in locations of low values
of $p(x,t)$, causing the bump to move away from locations of strong
synaptic depression. In turn, the region of synaptic depression tends
to follow the bump. However, if the time scale of synaptic depression
is large, the recovery of the synaptic depressed region is slowed
down, and cannot catch up with the bump motion. Thus, the bump starts
moving spontaneously.

To incorporate asymmetry into the moving state, we propose the following
ansatz:\begin{eqnarray}
    u(x,t) & = & u_{0}(t)\exp\left[-\frac{(x-vt)^{2}}{4a^{2}}\right],
\label{eq:u_moving}\\
    p(x,t) & = & 1-p_{0}(t)\exp\left[-\frac{(x-vt)^{2}}{2a^{2}}\right]
    +p_{1}(t)\exp\left[-\frac{(x-vt)^{2}}{2a^{2}}\right]
    \left(\frac{x-vt}{a}\right).
\label{eq:p_moving}
\end{eqnarray}
Projecting the terms in Eq. (\ref{eq:dyn}) to the basis functions
$\exp\left[-(x-vt)^{2}/(4a^{2})\right]$ and $\exp\left[-(x-vt)^{2}/(4a^{2})\right](x-vt)/a$,
and those in Eq. (\ref{eq:dp}) to $\exp\left[-(x-vt)^{2}/(2a^{2})\right]$
and $\exp\left[-(x-vt)^{2}/(2a^{2})\right](x-vt)/a$, we obtain four
equations for $\overline{u}$, $p_{0}$, $p_{1}$ and $v\tau_{s}/a$.
Real solutions exist only if \begin{equation}
    \frac{\overline{\beta}\overline{u}^{2}}
    {1+\overline{k}\overline{u}^{2}/8}\ge
    A\left[\frac{\tau_{d}}{\tau_{s}}-B
    +\sqrt{\left(\frac{\tau_{d}}{\tau_{s}}-B\right)^{2}-C}
    \right]^{-1},
\label{eq:moving}
\end{equation}
where $A=7\sqrt{7}/4$, $B=(7/4)[(5/2)\sqrt{7/6}-1]$,
and $C=(343/36)(1-\sqrt{6/7})$. As shown in Fig. 3, the boundary
of this region effectively coincides 
with the numerical solution of the line separating 
the static and moving phases.

Note that when $\tau_{d}/\tau_{s}$ increases, the static phase shrinks.
This is because the recovery of the synaptic depressed region is slowed
down, making it harder to catch up with changes in the bump motion.

\begin{figure}[ht]
\centering
\vspace{0pc}
\includegraphics[width=18pc]{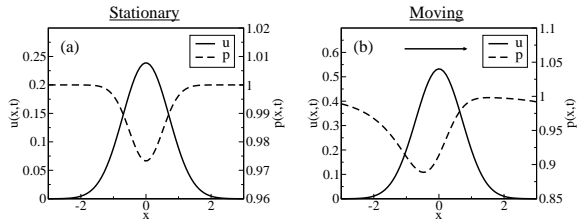}\hspace{0.5pc}%
\begin{minipage}[b]{13pc}\caption{
Neuronal input $u(x,t)$ and the STD coefficient $p(x,t)$ in
(a) the static state at $(\overline{k},\overline{\beta})=(0.9,0.005)$,
and (b) the moving state at $(\overline{k},\overline{\beta})=(0.5,0.015)$.
Parameter: $\tau_d/\tau_s=50$.
}
\end{minipage}
\end{figure}

An alternative approach that arrives at Eq. (\ref{eq:moving}) is to
consider the instability of the static bump, which is obtained by
setting $v$ and $p_{1}$ to zero in Eqs. (\ref{eq:u_moving}) and
(\ref{eq:p_moving}). Considering the instability of the static 
bump against the asymmetric fluctuations in $p_{1}$
and $vt$, we again arrive at Eq. (\ref{eq:moving}). This shows
that as soon as the moving bump comes into existence, the static bump
becomes unstable.
This also implies that in the entire region that the static and moving
bumps coexist, the static bump is unstable to asymmetric fluctuations.
It is stable (or more precisely, metastable) 
when it is static, 
but once it is pushed to one side, it will continue to move along that
direction. We may call this behavior \emph{metastatic}. As we shall
see, this metastatic behavior is also the cause of the enhanced tracking
performance.

\subsection{The Plateau Behavior}

To illustrate the plateau behavior, we select a point
in the marginally unstable regime of the silent phase,
that is, in the vicinity of the static phase. As shown in Fig. 8,
the nullclines of $u$ and $p_{0}$ ($d\overline{u}/dt=0$ and 
$dp_{0}/dt=0$ respectively)
do not have any intersections as they do in the static phase 
where the bump state exists. Yet,
they are still close enough to create a region with very slow dynamics
near the apex of the $u$-nullcline at
$(\overline{u},p_{0})=[(8/\overline{k})^{1/2},
\sqrt{7/4}(1-\sqrt{\overline{k}})]$.
Then, in Fig. 8, we plot the trajectories of the dynamics starting
from different initial conditions.
For verification, we also solve the full equations
(\ref{eq:dyn}) and (\ref{eq:dp}),
and plot a flow diagram with the axes being $\max_x u(x,t)$
and $1-\min_x p(x,t)$.
The resultant flow diagram has a satisfactory agreement with Fig. 8.

The most interesting family of trajectories
is represented by B and C in Fig. 8.
Due to the much faster dynamics of $\overline{u}$, trajectories starting
from a wide range of initial conditions converge rapidly, in a time
of the order $\tau_{s}$, to a common trajectory in the close neighborhood
of the $\overline{u}$-nullcline. Along this common trajectory, $\overline{u}$
is effectively the steady state solution of Eq. (\ref{eq:dU0_app})
at the instantaneous value of $p_{0}(t)$, which evolves with the
much longer time scale of $\tau_{d}$. This gives rise to the plateau
region of $\overline{u}$ which can survive for a duration of the
order $\tau_{d}$. The plateau ends after the trajectory has passed
the slow region near the apex of the $\overline{u}$-nullcline. 
This dynamics is in clear contrast with trajectory D, 
in which the bump height decays to zero 
in a time of the order $\tau_s$. 

Trajectory A represents another family of trajectories having
rather similar behaviors, although the lifetimes of their plateaus
are not so long. These trajectories start from more depleted initial
conditions, and hence do not have chances to get close to the 
$\overline{u}$-nullcline. 
Nevertheless, they converge rapidly, in a time of order
$\tau_{s}$, to the band $\overline{u}\approx(8/\overline{k})^{1/2}$,
where the dynamics of $\overline{u}$ is slow. The trajectories then
rely mainly on the dynamics of $p_{0}$ to carry them out of this
slow region, and hence plateaus of lifetimes of the order $\tau_{d}$
are created.

\begin{figure}[ht]
\centering
\vspace{3mm}
\begin{minipage}{17pc}
\centering
\includegraphics[width=10pc]{phase_plane_theory.eps}
\caption{
Trajectories of network dynamics starting from various initial conditions at
$(\overline{k}, \overline{\beta})$ = (0.95, 0.0085) (point P in Fig. 3). 
Solid line: $\overline{u}$-nullcline. 
Dashed line: $p_0$-nullcline. 
Symbols are data points spaced at time intervals of $2\tau_s$.}
\end{minipage}\hspace{1pc}%
\begin{minipage}{15pc}
\centering
\includegraphics[width=12pc]{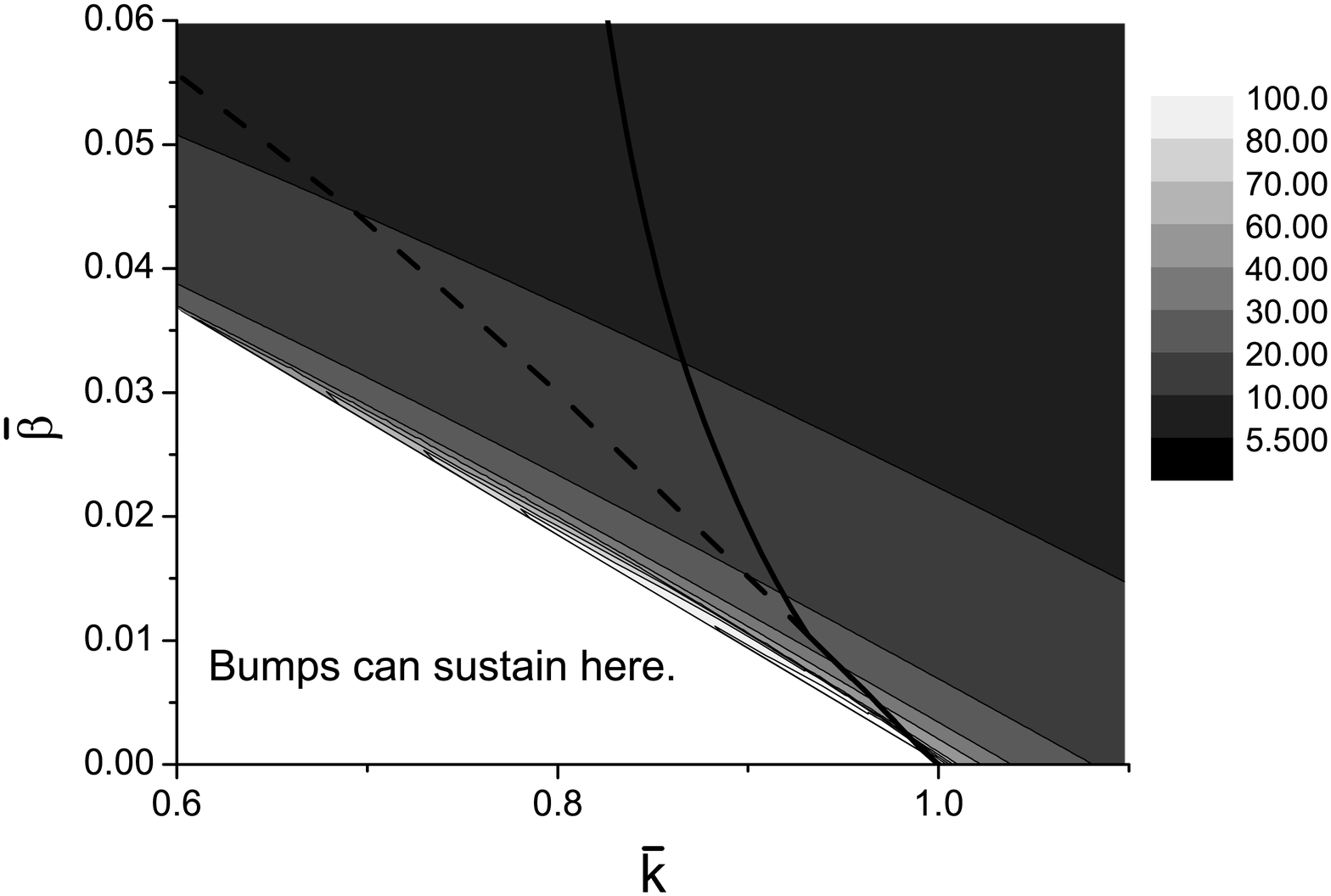}
\caption{
Contours of plateau lifetimes in the space of $\overline{k}$ and
$\overline{\beta}$.
The lines are the two topmost phase boundaries in Fig. 3. 
In the initial condition, $\alpha=0.5$.}
\end{minipage}
\end{figure}

Following similar arguments, the plateau behavior also exists in the
stable region of the static states. This happens when the initial
condition of the network lies outside the basin of attraction of the
static states, but it is still in the vicinity of the basin boundary.

When one goes deeper into the silent phase, 
the region of slow dynamics between the $\overline{u}$- 
and $p_0$-nullclines broadens. 
Hence plateau lifetimes are longest
near the phase boundary between the bump and silent states, and become
shorter when one goes deeper into the silent phase. 
This is confirmed by the contours of plateau lifetimes in the phase diagram 
shown in Fig. 9 obtained by numerical solution. 
The initial condition is uniformly set by introducing an external stimulus 
$I^{\rm ext}(x|z_0)=\alpha u_0\exp[-x^2/(4a^2)]$ 
to the right hand side of Eq. (\ref{eq:dyn}), 
where $\alpha$ is the stimulus strength. 
After the network has reached a steady state, 
the stimulus is removed at $t=0$, 
leaving the network to relax.

\subsection{The Tracking Behavior}

To study the tracking behavior, 
we add the external stimulus $I^{\textrm{ext }}(x|z_{0}) =\alpha
u_0\exp\left[-(x-z_{0})^{2}/(4a^{2})\right]$ 
to the right hand side of 
Eq. (\ref{eq:u_moving}), where $z_{0}$ is the position of the stimulus 
abruptly changed at $t=0$. 
With this additional term, we solve the modified version of
Eqs. (\ref{eq:u_moving}) and (\ref{eq:p_moving}),
and the solution reproduces the qualitative features due to the presence
of synaptic depression, namely,
the faster response at weak $\overline{\beta}$,
and the overshooting at stronger $\overline{\beta}$. As remarked
previously, this is due to the metastatic behavior of the bumps, which
enhances their reaction to move from the static state when a small
push is exerted.

However, when describing the overshooting of the tracking process,
the quantitative agreement between the numerical solution
and the ansatz in Eqs. (\ref{eq:u_moving}) and (\ref{eq:p_moving})
is not satisfactory.
We have made improvement
by developing a higher order perturbation analysis
using basis functions of the quantum harmonic oscillator~\cite{Fung10}.
As shown in Fig. 5, the quantitative agreement is much
more satisfactory.

\section{Conclusions and Discussions}

In this work, we have investigated the impact of STD on the dynamics of a CANN, and found that the network can support both static and moving
bumps. Static bumps exist only when the synaptic depression is sufficiently weak. A consequence of synaptic depression is that it places static
bumps in the metastatic state, so that its response to changing stimuli is speeded up, enhancing its tracking performance.  We conjecture that moving bump states may be associated with traveling wave behaviors widely observed in the neurocortex.

A finding in our work with possibly very important biological implications is that STD endows the network with slow-decaying behaviors. When
the network is initially stimulated to an active state by an external input, it will decay to silence very slowly after the input is removed.
The duration of the plateau is of the time scale of STD 
rather than neural signaling, and it provides a way for the network to hold the stimulus
information for up to hundreds of milliseconds, if the network operates in the parameter regime that the bumps are marginally unstable. This
property is, on the other hand, extremely difficult to be implemented in attractor networks without STD. In a CANN without STD, an active state
of the network decays to silence exponentially fast or persists forever, depending on the initial activity level of the network.
Indeed, how to shut off the activity of a CANN has been a challenging issue that received wide attention in theoretical neuroscience, with
solutions suggesting that a strong external input either in the form of inhibition or excitation must be applied (see, e.g., \cite{Gutkin01}).
Here, we show that STD provides a mechanism for closing down network activities naturally and in the desirable duration.

We have also analyzed the dynamics of CANNs with STD
using a Gaussian approximation of the bump.
It describes the phase diagram of the static and moving phases,
the plateau behavior,
and provides insights on the metastatic nature of the bumps
and its relation with the enhanced tracking performance.
In most cases, approximating $1-p(x,t)$ by a Gaussian profile
is already sufficient to produce qualitatively satisfactory results.
However, higher order perturbation analysis is required
to yield more accurate descriptions of results
such as the overshooting in the tracking process (Fig. 5).

Besides STD, there are other forms of STP that may be relevant to realizing short-term memory. Mongillo et al.~\cite{Mongillo08} have recently
proposed a very interesting idea for achieving working memory in the prefrontal cortex by utilizing the effect of short-term facilitation
(STF). Compared with STD, STF has the opposite effect in modifying the neuronal connection weights. The underlying biophysics of STF is the
increased level of residual calcium due to neural firing, which increases the releasing probability of neural transmitters. Mongillo et
al.~\cite{Mongillo08} showed that STF provides a way for the network to encode the information of external inputs in the facilitated connection
weights, and it has the advantage of not having to recruit persistent neural firing and hence is economically efficient. 
This STF-based memory mechanism is, however, 
not necessarily contradictory to the STD-based one we propose here. 
They may be present in different cortical areas 
for different computational purposes. 
STD and STF have been observed to have different effects 
in different cortical areas. 
One location is the sensory cortex 
where CANN models are often applicable. 
Here, the effects of STD tends to be stronger than that of STF. 
Different from the STF-based mechanism, 
our work suggests that the STD-based one 
exhibits the prolonged neural firing, 
which has been observed in some cortical areas. 
In terms of information transmission, 
prolonged neural firing is preferable in the early information pathways, 
so that the stimulus information can be conveyed to higher cortical areas
through neuronal interactions. 
Hence, it seems that the brain may use a strategy of weighting the effects of STD and STF differentially for carrying
out different computational tasks. It is our goal in future work to explore the joint impact of STD and STF on the dynamics of neuronal
networks.

This work is partially supported by the Research Grants Council of Hong
Kong (grant nos. HKUST 603607 and 604008).

\end{document}